\documentclass[preprint]{aastex}

\usepackage{graphicx}
\usepackage{natbib}
\def\icarus{Icarus}

\def\Soh{{\Sigma_{_0}}}
\def\hoh{{h_{_{0}}}}
\def\aoh{{a_{_{0}}}}
\def\ain{{a_{_{\rm in}}}}
\def\aout{{a_{_{\rm out}}}}
\def\orchestra{\textit{Orchestra}}
\def\XX{\tilde{X}}

\newcommand{\lae}{\lower 2pt \hbox{$\, \buildrel {\scriptstyle <}\over {\scriptstyle\sim}\,$}}
\newcommand{\gae}{\lower 2pt \hbox{$\, \buildrel {\scriptstyle >}\over {\scriptstyle\sim}\,$}}

\def\eqnum[#1]{(\ref{#1})}

\def\AU{{\rm AU}}
\def\Msolar{{\rm M}_\odot}
\def\Mearth{{\rm M}_\oplus}

\def\mearth{\ifmmode {\rm M_{\oplus}}\else $\rm M_{\oplus}$\fi}
\def\msun{\ifmmode {\rm M_{\odot}}\else $\rm M_{\odot}$\fi}

\def\Mstar{M_\star}

\def\mcrit{m_{_{\rm crit}}}

\def\vecdv{\delta \vec{v}}

\def\rp{r_\bullet}

\def\Caero{C_{_{\rm aero}}}
\def\Cdyn{C_{_{\rm dyn}}}
\def\Cgrav{C_{_{\rm grav}}}
\def\mp{m_\bullet}

\def\rhog{\rho_{_{\rm gas}}}
\def\cs{c_s}
\def\adyn{a_{_{\rm dyn}}}
\def\aaero{a_{_{\rm aero}}}

\def\ma{\mu}

\def\vkep{v_{_{\rm Kep}}}

\def\openrate{{\kappa_{_{\rm in}}}}
\def\hole{cavity}

\begin{document}

\title{The fate of scattered planets}

\author{Benjamin C. Bromley}
\affil{Department of Physics \& Astronomy, University of Utah, 
\\ 115 S 1400 E, Rm 201, Salt Lake City, UT 84112}
\email{bromley@physics.utah.edu}
\author{Scott J. Kenyon}
\affil{Smithsonian Astrophysical Observatory,
\\ 60 Garden St., Cambridge, MA 02138}
\email{skenyon@cfa.harvard.edu}

\begin{abstract}

As gas giant planets evolve, they may scatter other planets far from
their original orbits to produce hot Jupiters or rogue planets that
are not gravitationally bound to any star.  Here, we consider planets
cast out to large orbital distances on eccentric, bound
orbits through a gaseous disk.  With simple numerical models, we show 
that super-Earths can interact with the gas through dynamical friction 
to settle in the remote outer regions of a planetary system. Outcomes
depend on planet mass, the initial scattered orbit, and the evolution
of the time-dependent disk. Efficient orbital damping by dynamical
friction requires planets at least as massive as the Earth.
More massive, longer-lived disks damp eccentricities more efficiently
than less massive, short-lived ones. Transition disks with an
expanding inner \hole\ can circularize orbits at larger distances than
disks that experience a global (homologous) decay in surface density. 
Thus, orbits of remote planets may reveal the evolutionary history of 
their primordial gas disks. A remote planet with an orbital distance
$\sim$100~AU from the Sun is plausible and might explain correlations
in the orbital parameters of several distant trans-Neptunian objects.

\end{abstract}


\maketitle

\section{Introduction}

The formation of gas giants is a fast process limited by the lifetime
of the gas in a circumstellar disk.  In core accretion models, solids
coalesce rapidly into $\sim$10$\ \Mearth$ planets, which then accrete
gas to become full-fledged giants \citep{pol96}.  The core formation
step is uncertain.  A plausible mechanism involves large (1-1000~km)
seed objects that form quickly by gravitational streaming
instabilities \citep{joh07,dit13} and then accrete pebbles and
collision fragments which interact aerodynamically with the gas
\citep{kb09,kob11,bk11a,lam12,cha14}.
Once a few large cores form, competition for the remaining fragments
and pebbles favors the most massive protoplanets.  As gas giants begin to 
carve out gaps in the disk, numerical simulations show that gas giants
gravitationally scatter other massive cores, leftover planetesimals, 
and other less massive objects \citep[e.g.,][]{cha08,bk11a}.  

Scattering appears to play a major role in setting the orbital 
architecture of many planetary systems.  Dynamical interactions between 
gas giants and leftover large planetesimals stabilized the outer solar
system \citep[e.g.,][]{tsi05,morbi2013}. Among exoplanets, scattering 
can explain the high orbital eccentricities of ice and gas giants close 
to the host star \citep[e.g.,][]{jur08,for08,cha08}.  At the other 
extreme, free-floating planets can result from ejection during a strong 
gravitational encounter with a more massive planetary companion
\citep{lev98}. Weaker encounters result in planets which are scattered 
outward but remain bound to the host star \citep[e.g.,][Figs.~14--16]{bk11a}.

With initial trajectories resembling the orbits of long-period comets, 
the ultimate fate of planets on eccentric orbits far from their host stars
depends on their interactions with the gas and solids remaining in the 
protoplanetary disk. In a low mass disk, weak interactions probably prevent 
the orbit from circularizing, leaving the planet on an eccentric orbit and
risking additional interactions with more massive planets closer to the host 
star. If circularization is possible in a more massive disk, the scattered
planet may find a stable orbit far from its birthplace.

Quantifying the probability of scattering and subsequent
circularization of giant planets has clear observational
implications. Among the known exoplanets, the 1.5~$\Msolar$ star
HR~8799 has a planetary system with four super-Jupiters on low
eccentricity orbits with semimajor axes of roughly 15--70~AU
\citep{mar08,cur12b}.  Although migration from 5--10~AU is a popular
model for producing at least some of the observed planets
\citep[e.g.,][]{hah99,cri09,ray14}, scattering followed by
circularization is a plausible alternative.  With a highly eccentric
orbit ($e \sim$ 0.8) far from its host star ($a \approx$ 120~AU),
Fomalhaut~b is a promising candidate for a low mass planet scattered
during a strong gravitational encounter with a more massive planet
\citep{kal08,kal13,beu14,tamo14}. However, the mass of the planet and
the orbital eccentricity remain very uncertain
\citep[e.g.,][]{cur12a,kcb14}.  As direct imaging reveals larger
samples of exoplanets on wide orbits \citep{mac14,tamu14}, robust
constraints on the physical conditions required for scattering and
circularization will enable better evaluations of plausible formation
mechanisms for these systems.

Here, we describe outcomes for planets scattered into remote regions
of their planetary systems during the epoch of gas giant formation.
Our main goal is to highlight physical conditions in a protostellar
disk which enable a scattered planet to settle into a low-eccentricity
orbit at a large distance from the planet forming region. In
\S\ref{sect:theory} we outline the structure and evolution of a disk,
along with a prescription for planet-disk interaction by gas drag and
dynamical friction. Then, in \S\ref{sect:sims} we present results from
simulations of scattered planets as they interact with an evolving disk.
Finally, in \S\ref{sect:discuss} we summarize our results and put them
in context with observations of exoplanetary systems.

\section{Scattered planets and protoplanetary disks: preliminaries}
\label{sect:theory}

To calculate the long-term evolution of interactions between scattered 
planets and gas disks, we use the N-body component of \orchestra, an ensemble
of computer codes for the formation and evolution of planetary systems
\citep{bk06,kb09,bk11a}. This code, with an adaptive 6$^{\rm th}$ order
integrator, has passed a stringent set of dynamical tests and benchmarks
\citep[e.g.,][]{dun98}. We have used the code to simulate scattering of
super-Earths by growing gas giants \citep{bk11a}, migration through
planetesimal disks \citep{bk11b} and Saturn's rings \citep{bk13},
and formation of Pluto's small satellites \citep{kb14a}.

Here, we modify the code to include dynamical friction and aerodynamic
drag from the gas disk using analytical approximations for the
acceleration \citep[e.g.,][]{ost99}.  We then consider the orbits of
individual planets scattered onto distant, highly eccentric orbits by
one or more larger planets orbiting at smaller distances from the host
star.  We do not attempt to model planet-planet scattering in detail.
Instead, we describe general conditions which lead to a remote planet
on an eccentric orbit, dynamically isolated from the rest of the
planetary system except for the extended gas disk
(\S\ref{sect:scatter}). We then describe our parameterization of the
gas disk and its evolution (\S\ref{sect:disk}), the acceleration on a
scattered planet by the disk (\S\ref{sect:interx}), and the results of
simulations (\S\ref{sect:sims}).

\subsection{Growing cores and planet-planet scattering}
\label{sect:scatter}
 
Our scenario for creating a remote planet has a key preliminary step,
the formation of multiple planetary cores before the gas disk
disperses. Within this time frame, at least one of these cores must
reach a critical mass, $\mcrit$, sufficient to scatter one or more of 
the other cores to large orbital distance.

We  crudely estimate $\mcrit$ for a core by considering an idealized 
encounter with a lighter companion.  We assume that the larger planet 
is on a circular orbit. The smaller planet, through previous interactions
with other planets, has negligible orbital speed, as if near apoastron
on an eccentric orbit.  We also assume that the interaction sends the
smaller planet radially outward from the host star. Thus, it experiences
a $90^\circ$ deflection in the reference frame of the larger planet.
The critical mass then follows from a Rutherford scattering analysis
with the constraint that the smaller core must have a distance of
closest approach that remains outside of the physical radius of the
larger one:
\begin{equation}
\mcrit \approx 
  10
  \left(\frac{a}{5\ \AU}\right)^{\!-3/2}
  \left(\frac{\rho}{2\ \mbox{g/cm}^3}\right)^{\!-1/2}
  \left(\frac{\Mstar}{1\ \Msolar}\right)^{\!-3/2}
  \ \Mearth
\end{equation}
where $a$ is the planet's orbital distance, $\rho$ is its average mass
density \citep[$\rho \approx 2$~g/cm$^3$ corresponds to a core with a
  mixture of ice and rock; see][]{pap99}, and $\Mstar$ is the mass of
the central star.  In principle, significant outward scattering can
occur even as the larger core is just starting to accrete gas from the
protostellar disk \citep{pol96,dan11,hor11,pis14}.  The formation time
for a core of this mass is uncertain, but simulations suggest that gas
accreting cores can form well within the lifetime of the gas disk
\citep[see, for example][]{kb09,bk11a}.

Numerical simulations of gas giant formation confirm that 
(a) multiple cores can form, and (b) they can scatter each other to 
large distances \citep{bk11a}. In typical models that produce multiple 
planets, over 80\% of the 1--15~$\Mearth$ cores get scattered beyond 
$\sim 30$~AU. Although the timing is uncertain, scattered planets are 
probably a common outcome of gas giant planet formation.

After scattering by a more massive core, a remote planet's orbit 
continues to evolve as it repeatedly encounters the same massive
core every periastron passage. There are several reasons to suspect 
that repeated encounters following a large scattering event do not 
greatly alter the remote planet's orbit. First, the time between 
periastron passages, i.e., the remote planet's orbital period, is 
\begin{equation}
T_{\rm orbital} = 1000 \left(\frac{a}{100~{\rm AU}}\right)^{3/2}
\left(\frac{\Mstar}{\Msolar}\right)^{-1/2} \ {\rm yr} .
\end{equation} 
This time scale is only somewhat shorter than the time scale for
the massive core with mass $m$ to migrate through the disk
\citep{war97,tan02,pap07}: 
\begin{equation}\label{eq:mig}
T_{\rm migrate} \equiv \frac{a}{\dot{a}} \approx 3\times 10^4
\left(\frac{m}{10~\Mearth}\right)^{-3/2}
\left(\frac{a}{5~{\rm AU}}\right)^{1/2}
\left(\frac{\Mstar}{\Msolar}\right)^{3/2} \ {\rm yr}
\end{equation}
Thus, between each periastron passage of the remote planet, a 
$10\ \Mearth$ core at 5~AU moves by roughly 0.2~AU. This 
distance is slightly larger than the massive core's Hill radius, 
which defines its gravitational sphere of influence as it orbits 
the host star. Thus, unless the two planets have a rare orbital
commensurability, the larger planet may drift inward, leaving the 
orbit of the remote planet free from subsequent interactions. 

Other mechanisms, such as gravitational perturbations from other
planetary cores or nearby stars (if the young host star is in a
cluster), may serve to isolate the remote planet from the massive 
core that scattered it. Interactions between the remote planet and 
the gas disk may change the planet's orbit and further isolate it
from the massive core. To estimate the time scale for these 
interactions, we now consider the properties of the gas disk.

\subsection{Disk structure and evolution}
\label{sect:disk}

Once flung outward, the smaller planet interacts with the gas and 
solid particles in the outer disk.  To assess the effect of the disk 
on a scattered planet's orbit, we establish the physical properties of
circumstellar disks. At early times, disks are massive and extended.
For solar-type stars with typical ages of 1--2~Myr, total disk masses
are $0.001$--0.1~$\Msolar$ \citep{and13}.  Disks evolve on a time scale
$\tau$ of several million years \citep{hai01} through viscous
dissipation \citep{lyn74,lin80,lin82}, photoevaporation by the
radiation of the host star \citep{cla01,owe12}, and erosion from
stellar winds \citep{rud04,lov08}.  Together, these interactions lead 
to a general dispersal of the disk, with a monotonic decrease in surface 
density over time at rates which may depend on orbital distance from the 
host star \citep[as in the reviews by][]{ale13,you13}.

We assume the gaseous component of the disk has an axisymmetric 
surface density distribution as a function of orbital distance $a$ 
and time $t$: 
\begin{equation}\label{eq:disk}
\Sigma(a,t) = 
\Soh X e^{-t/\tau} 
\left(\frac{a}{\aoh}\right)^{\!-p}  ,
\end{equation}
where the power-law index $p \sim 1$, $\Soh \equiv
2000$~g/cm$^2$ is a fiducial surface density at distance $\aoh \equiv
1$~AU, and the parameter $X$ scales the initial mass of the disk
\citep[cf.][]{pas04,pie07,and11,den13,bir14}.

The disk has a vertical scale height 
\begin{equation}\label{eq:height}
H(a) = \hoh \left(\frac{a}{\aoh}\right)^{q},
\end{equation}
where $\hoh/\aoh = 0.01$--0.05 and the power-law index $q = 9/7$
\citep{cha97}. This scale height is proportional to the sound speed in
the gas, $\cs \approx H\vkep/a$, where $\vkep$ is the circular Keplerian
speed at orbital distance $a$.  We assume that $H$ and $\cs$ are
independent of time. 

The gas density within the disk is approximately $\Sigma/H$:
\begin{eqnarray} 
\label{eq:rhog}
\rhog & = & \frac{\Soh X}{\hoh} e^{-t/\tau} 
              \left(\frac{a}{\aoh}\right)^{-p-q} \\ 
\ & \approx & 4.5 \times 10^{-9}\, \XX \, e^{-t/\tau} \,
 \left(\frac{a}{1 \AU}\right)^{\!16/7} \ \mbox{g/cm$^3$}\ 
\ \ \ \ (p = 1).\rule{0pt}{22pt}
\end{eqnarray} 
where 
\begin{equation}\label{eq:XX} 
\XX = X \frac{0.03}{\hoh/\aoh} .
\end{equation}
Introducing $\XX$ allows us to parameterize the gas density, 
which regulates where and when a scattered planet settles in 
the outer regions of a planetary system.

The final property is the rotation speed of the gaseous disk. We assume 
material in roughly circular orbits set by the stellar and disk potentials.
Gas pressure reduces the orbital velocity by a factor of $(1-H^2/a^2)$ 
\citep[e.g.,][]{you13}.

To model the time evolution of the disk, we consider two prescriptions
for the monotonic decline in disk surface density.  To describe the global
loss of gas from viscous processes, equation~(\ref{eq:disk}) includes 
a term which allows us to set the exponential loss of gas on time scales 
of $\tau = 1$--10~Myr \citep[cf.][]{hai01}. 

To consider the possibility
of the inside-out decay that produces a transitional disk, we establish
an inner edge at orbital distance $\ain$ that linearly expands with time.
Defining a constant expansion rate, $\openrate$, the inner edge evolves 
as:
\begin{equation}
  \ain(t) = \ain(0) + \openrate t ~ ,
\end{equation}
until the inner edge reaches the fixed outer edge $\aout$. Once $\ain$ =
$\aout$, the disk mass reaches zero; interactions between the planet and
the gas cease. Numerical
simulations \citep[e.g.,][]{owe12} and observations of transition disks
\citep{cal05,cur08,and11} suggest opening rates of roughly 10~AU/Myr.  

If a gas giant starts the dispersal by carving out a gap in the disk,
the inside-out removal of disk material is plausible 
\citep[see][and references therein]{ale13}.  Here, we always assume that
a massive gas giant orbiting at roughly 5~AU is responsible for scattering
planets into the outer disk.  Although the exponential decay of a disk 
from viscous evolution and the erosion of the disk inner edge by 
photoevaporation and stellar wind erosion probably occur simultaneously
\citep[cf.][]{rib14}, here we consider these modes as separate cases.

\subsection{Planet-disk interactions}
\label{sect:interx}

Planets interact with the gas aerodynamically and gravitationally.  
Gas flowing by the planet produces aerodynamic drag. The gravity of 
planets with $M \gtrsim 1\ \Mearth$ can create a density wake in 
the disk; the dynamical friction associated with this wake tends to
circularize the planet's orbit. For both mechanisms, the amount of
drag depends on the mass and radius of the planet and the mean density, 
the sound speed, and other properties of the disk
\citep{dok64,rud71,wei77,oht88,ost99,adac09,lee14}.

To estimate the net acceleration from these two processes, we calculate
\begin{equation}\label{eq:acc}
  \frac{d\vec{v}_{_{\rm drag}}}{dt} = 
  -\max(\aaero,\adyn)\frac{\vecdv}{|\vecdv|}
\end{equation}
where $\vecdv$ is the velocity of the planet in the rest frame of the 
surrounding gas, and 
\begin{eqnarray}\label{eq:drag}
  \adyn & = & 
    \frac{2\pi G^2\rhog\mp}{\cs^2}
    \frac{\ma^2(\Caero^2+\Cdyn^2\ma^2)^{1/2}}{(1+\ma^2)^{5/2}}
\\ \label{eq:aero}
  \aaero & = & 
\frac{\pi\Caero\rp^2\rhog|\vecdv|^2}{2\mp}
\end{eqnarray}
are associated with dynamical friction and aerodynamic drag
respectively; $\mp$ is the planet mass, $\rp$ is its physical radius,
$\ma \equiv |\vecdv|/\cs$ is the Mach number, and the $C$'s are drag
coefficients of order unity. We choose the form of the expression for
$\adyn$ to give desired results in the low- and high-Mach number
regimes, with an interpolation function to cover transonic speeds
in the manner of \citet{lee14}. Numerical simulations \citep[e.g.][]{ruf96}
indicate that this type of parameterization is realistic.

We estimate numerical values for the coefficients $\Caero$ and $\Cdyn$
with simple assumptions. In the aerodynamic case, the planet is much
larger than the mean free path in the gas. For subsonic speeds, the 
frictional acceleration is then proportional to $|\vecdv|^2$. A 
spherical shape in this (quadratic) regime corresponds to $\Caero = 0.44$
\citep[e.g.][]{ada76,wei77}. 

For supersonic flow, we estimate the magnitude of the dynamical
friction force from integrating over the impact parameter $b$ for gas
elements as they flow past the planet.  Weak scattering theory gives
the contribution from each element to the acceleration as $\delta a
\sum (b|\vecdv|)^{-2}$ \citep[e.g.,][]{dok64,rud71,lin79}.  We
consider only more distant encounters with $b>H/2$; on scales smaller
than the disk height, random motions of the gas wash out the effect.
Then, assuming that the planet is traveling near the midplane of a
disk with slab geometry, we integrate over all gas streamlines flowing
by, except for those streamlines that come within a distance $H/2$ of
the planet in the disk plane\footnote{Integration over $\delta a$ from
  gas streamlines that pass as close as the surface of the planet
  yield a coefficient $\Cdyn$ with approximate
  logarithmic dependence on physical radius, as in a Coulomb integral
  \citep{bintre87}.  Here, we exclude streamlines flowing through a
  square region of dimensions $H\times H$, centered on the planet, as
  well as any streamlines that lie above or below the slab with
  elevation $|z|>H/2$, where $\rhog = 0$. In the limit of an extended
  slab, the result is a constant, independent of $H$.  In excluding
  the contribution from gas flowing near the planet, we 
  underestimate the strength of dynamical friction.}.  Following
this prescription, we estimate $\Cdyn = 0.62$.

In addition to the small-scale gravitational wakes, planets orbiting 
within the disk are accelerated by long-range interactions with the 
full disk.  Assuming that the disk on large scales is largely 
unperturbed by the planet, the disk potential is
axisymmetric, determined by disk parameters $\ain$, $\aout$, $\hoh$,
and $\Sigma$. For example, in a geometrically-thin power-law disk,
the acceleration at orbital distance $a$ near the midplane is
\begin{equation}\label{eq:diskgrav}
  \vec{a}_{_{\rm disk}} = -2\pi \Cgrav G \Sigma(r) \frac{\vec{r}}{r} \ ,
\end{equation}
in the limit $\ain \ll a \ll \aout$, where $\Cgrav$ is a constant of
order unity that depends on the power-law index $p$
\citep[Appendix~A]{bk11a}.  

In practice, we calculate the unperturbed disk acceleration using a 
model with constant disk thickness of $\hoh$, and an efficient 
numerical algorithm that can accommodate an arbitrary surface density 
profile \citep{bk11a}.

\subsection{Limitations of the model}

Our approach to the acceleration includes the major large-scale forces,
drag and dynamical friction, between the planet and the disk. However,
we neglect the differential torques
between the planet and disk which generate type I and type II migration
\citep{war97}. Our goal instead is to follow the eccentricity damping
prior to orbit circularization. The physics in this case depends on
the bulk flow of gas, not on the gradient of the flow from Keplerian 
shear which is central to migration theory \citep{lin79,gt79}.  While 
migration can be important once the orbit circularizes (eq.~\ref{eq:mig}), 
the direction and pace depend on the thermodynamic state of the disk 
\citep{paa09}. To isolate the important issues involved in eccentricity
damping, we consider the thermodynamic state fixed and save detailed 
calculations for a future study.

We also ignore the possibility of additional accretion onto scattered
planets as they interact with gas in the outer regions of the disk. If
planets accrete \citep{pol96,raf11}, they will interact more strongly 
with the gas, enhancing the effects considered here.  Calculations 
suggest that growth from gas accretion may be important, even for 
planets as low as a few Earth masses if they are located beyond 100~AU 
\citep{pis14}. Thus the orbital damping times in this work are overestimates.

The scenario we propose here and the approximations introduced above
have several additional limitations. One concern is the
parameterization of the drag force (eq.~[\ref{eq:acc}]), particularly
in the transonic and subsonic regimes. Fortunately, by the time a
planet reaches subsonic speeds, it is already settled in the disk with
a relatively low eccentricity, $e \approx H/a \ll 1$. The supersonic
regime has more firm analytical basis \citep[e.g.,][although see Lee
  \& Stahler 2014]{ost99} that is consistent with numerical
simulations \citep{ruf96}. Furthermore, our implementation is
conservative in its neglect of interactions between the planet and gas
on scales less than the disk height; at supersonic speeds, we might
have included streamlines that crossed inside the planet's Hill
sphere.

We also limit the initial orbits of scattered planets to lie in the
disk midplane where planet-disk interactions are strongest.  When a
planet has a significant orbital inclination $i \gae H/a$, it 
spends only a fraction of its orbit embedded in gas. The orbital 
evolution is then similar to a system where a planet with small $i$
orbits within a disk with a lower gas surface density, with details 
that depend on the planet's orbital elements. Drawing from experience 
with dynamics in Saturn's rings \citep{bk13}, we speculate that when 
a planet starts out with some inclination, it will settle vertically
into the disk plane on a time scale similar to the circularization time.

In our calculations, we assume that the formation time for planets is
short compared to the evolutionary time scale of the disk. Thus,
fully-formed planets scatter at $t = 0$, when the disk has its initial 
surface density (eq.~[\ref{eq:disk}]).  Although core formation can be 
fast, well within $\sim$1~Myr \citep{kb09,lam12,pis14}, the surface 
density of the disk probably evolves as planets form 
\citep[e.g.,][]{bk11a}. This aspect of our approach tends to 
overestimate the disk mass available to circularize the planet and
underestimate the damping time. For the $\tau \approx$ 2--4~Myr disk
evolution times adopted here, the impact of this assumption is 
reasonably small.

\section{Simulations of planetary relocation}
\label{sect:sims}

To assess the effect of the disk on a scattered planet we set up a
suite of simulations with a variety of disk parameters and planetary
orbits around a 1~$\Msolar$ star. We use disk models with $\Sigma\sim
1/a$ ($p=1$), $\hoh/a = 0.03$, and a density factor $\XX$ with values
of 0.25, 0.5, and 1 (eqs.~[\ref{eq:disk}]--[\ref{eq:XX}]).  
%% @@@@ modified:
In all cases illustrated in the Figures, the disk starts with $\ain =
20$~AU and $\aout = 200$~AU.  
Simulations with $\ain = 5~AU$
and $\aout=350$~AU allow us to assess how outcomes depend on the initial
extent of the disk.
%the inner disk edge 
%inside of periastron and the outer edge beyond the orbit of the
%planet.  
To quantify the subsequent gas loss from the disk, we choose
either the exponential decay mode with $\tau = 2$ and 4~Myr, or an
expanding inner \hole\ with $\openrate = 20$ and 40~AU/Myr.

The models have initial disk mass that ranges from 0.06~$\Msolar$ to
0.25~$\Msolar$.  These values are large compared to those of disks
around T Tauri stars \citep{and13} but more in line with disks around
protostars \citep{and05}. This trend with stellar age is consistent
with the assumption that planets form rapidly 
\citep[see the discussion in][]{naj14}.  Still, the most massive
disk is probably somewhat unrealistic in terms of surface density and 
total mass. However, since the key physical quantity for planet-disk
interaction is the local density, the models labeled here according
to surface density can represent disks with lower (higher) total mass
if they have proportionally lower (higher) scale heights.

In our models, scattered planets have (i) masses of 1, 3, 10 and
30~$\Mearth$, (ii) initial periastron distances of 10~AU, and (iii)
initial apoastron distances of 100, 200 or 300~AU, corresponding to
$a$ of 55~, 105~AU and 155~AU, with $e$ of 0.82, 0.9 and 0.94,
respectively. For each configuration we ran simulations with a Jupiter
mass gas giant ($a=5$~AU, $e=0$); it had negligible effect on
outcomes. The orbits of all planets considered here are coplanar with
the disk.  In determining orbital elements we incorporate only the
central star's mass, without treating the gravity of the extended disk
or the gas giant. We launch all planets from periastron in the plane
of the disk and calculate evolution over a 10~Gyr period with a
6$^{\rm th}$-order integration code \citep{bk06,bk11a}.  With the most
massive planets in the most massive disks, we omit the orbital
evolution after the planet circularizes ($e<0.01$), as our code is not
designed to follow migration in this regime.

\subsection{Orbital evolution during disk depletion}

\begin{figure}[htb]
\centerline{\includegraphics[width=7.0in]{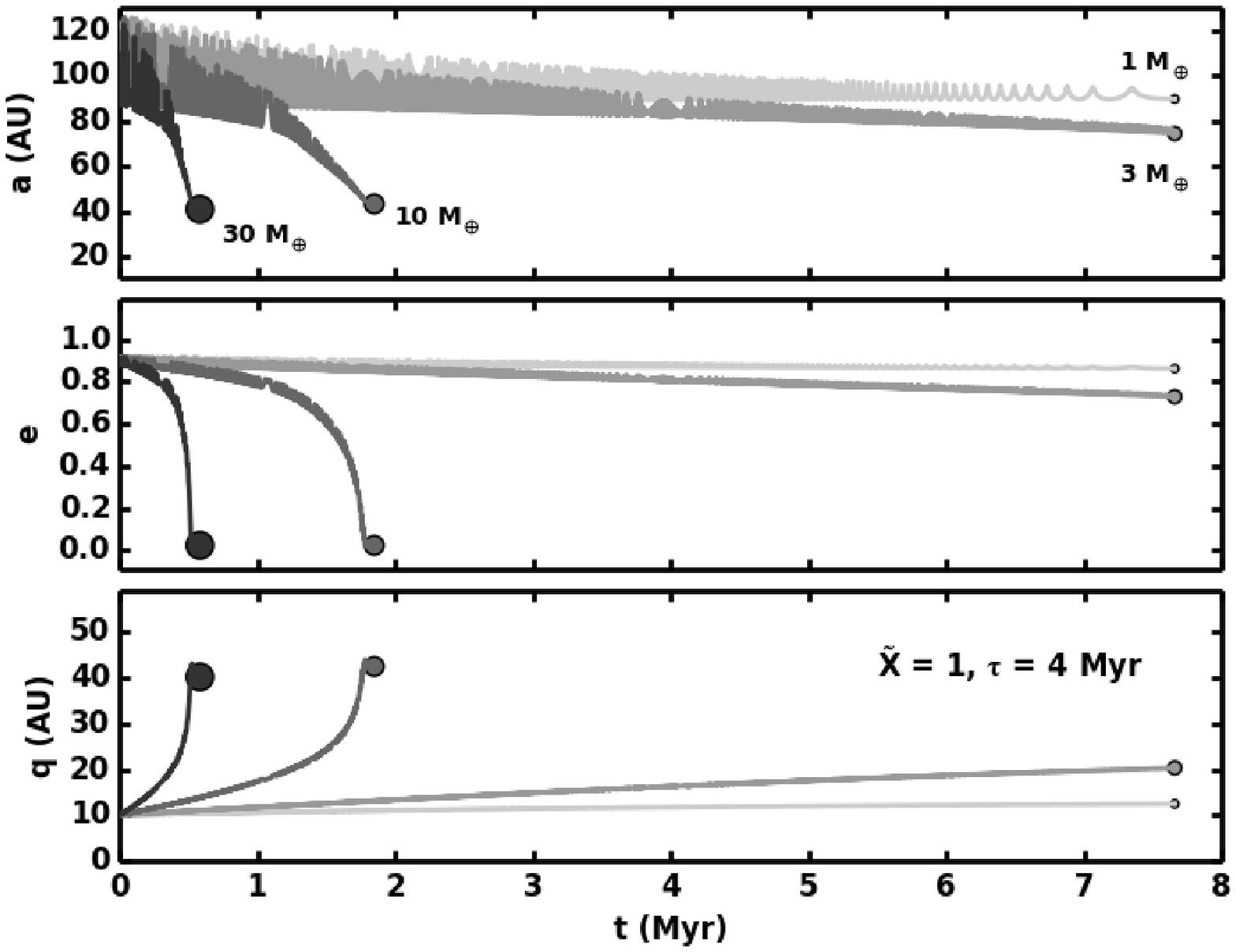}}
\caption{\label{fig:orbmp} Evolution of planetary orbits in an
  exponentially decaying disk as a function of planet mass.  Each
  track follows the evolution of a planet after it was placed on an
  orbit with an apoastron distance of 200~AU, in a
  disk with density parameter $\XX = 1$ and decay time scale of $\tau
  = 4$~Myr. Planet masses are labeled in the upper panel. In all
  panels, the darker shade tracks correspond to the more massive
  planets.  The trend illustrates the correlation of eccentricity
  damping with planet mass as a result of dynamical friction: a more
  massive planet can create a bigger density wake, which in turn has a
  stronger effect on its orbit. The high variability in the orbital
  elements, particularly the semimajor axis (upper panel), is an
  artifact of our estimators, which measure osculating parameters as
  if the disk were massless.  }
\end{figure}

To follow the impact of planet-disk interactions, we track the planetary 
orbital elements throughout each simulation.  Figure \ref{fig:orbmp} 
illustrates the orbital evolution of planets with masses ranging from
1--30~$\Mearth$ in a massive ($\XX=1$) disk that decays exponentially
with $\tau = 4$~Myr.  The more massive planets experience significant
circularization within a few million years; less massive planets remain 
on eccentric orbits with $e>0.5$. This strong dependence on planet mass 
is a direct result of dynamical friction, which generates an acceleration 
that scales as $\mp$ (eq.~[\ref{eq:drag}]). As in type I migration, a 
more massive planet creates a more massive density wake, which feeds 
back to affect the planet's orbit. Aerodynamic drag is not important for 
any of these planets over the disk lifetime.

The jitter in $a$, $e$, and $q$ in Figure \ref{fig:orbmp} is an
artifact of our approach to deriving orbital elements. To estimate
them, we assume osculating orbits in the Keplerian potential of the
star. However, at early times when the disk is massive, the osculating
orbital elements are affected by perturbations from the disk
potential.  Furthermore, when elements are sampled as a planet makes
radial excursions through the disk, the values of $a$, $e$, and $q$
can vary, even over a single orbital period. The result is a modest
amount of jitter. Because our goal is to follow trends in elements,
this choice for estimating orbital parameters has no impact on our
conclusions.

\begin{figure}[htb]
\centerline{\includegraphics[width=7.0in]{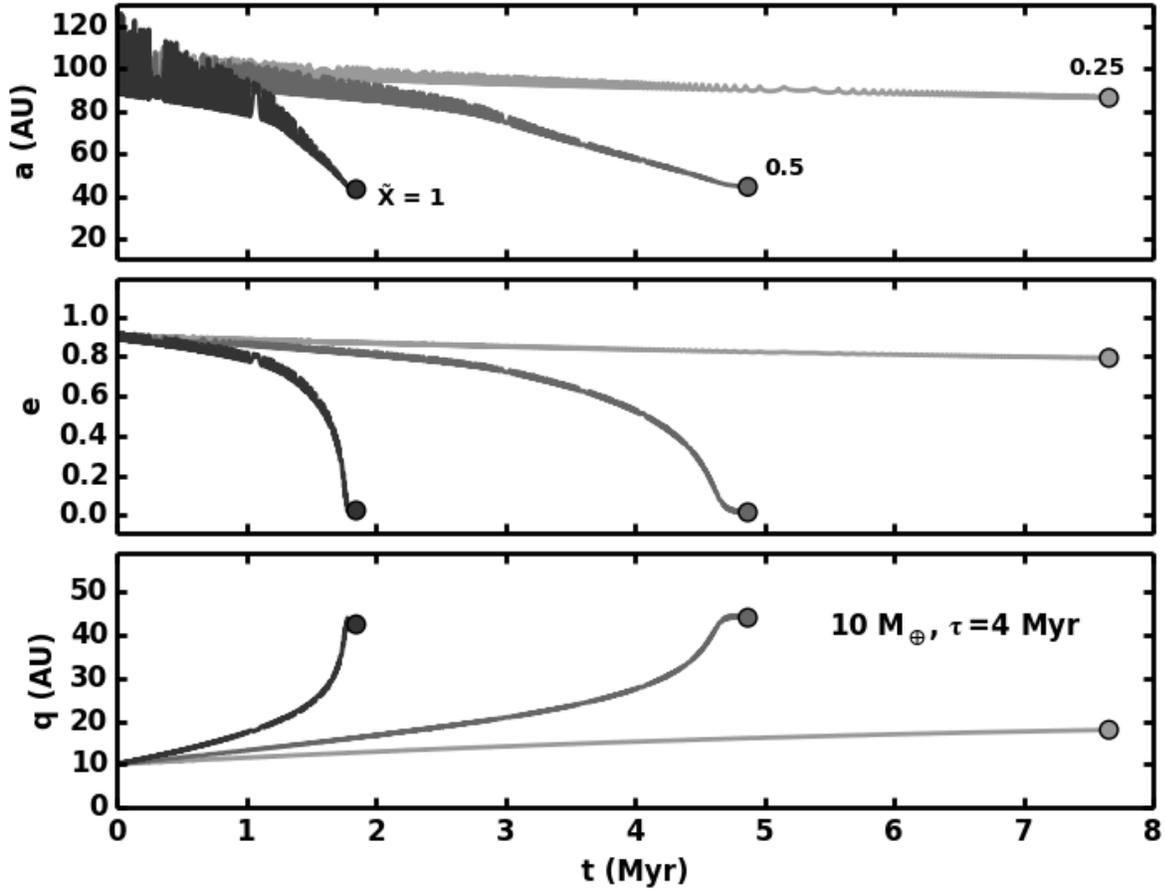}}
\caption{\label{fig:orbX} Evolution of orbits in disks with different
  density parameters. Orbits start off as in the previous figure, but
  in this case a $10\ \Mearth$ planet lies in disks with $\XX = 0.25$,
  0.5, and 1, as labeled in the upper panel. Darker shade of the lines
  and symbols indicates disks with higher density. The effect of the
  gas density is clear: higher $\rhog$ means a planet can create a
  more significant gravitational wake, hence it experiences stronger
  eccentricity damping.  
}
\end{figure}

To test the dependence of orbital evolution on disk structure, we 
consider the evolution of a planet with $\mp = 10$~$\Mearth$ in a
disk with gas density parameter $\XX$ = 0.25, 0.5 and 1
(Figure \ref{fig:orbX}). Here we see the clear impact of gas drag: 
higher gas density (larger $\XX$) causes stronger eccentricity 
damping. Massive planets in disks with $\XX$ = 1 damp on time scales
of 1--2~Myr. In less dense disks with $\XX$ = 0.25, there is little
damping after 7--8~Myr. 

In our models, the acceleration of the planet from dynamical friction
depends on the product of gas density and planet mass. Thus we might
be tempted to take advantage of this degeneracy and calculate a single
suite of simulations as a function of one parameter, $\gamma =
\mp\rhog$. Then we could estimate the behavior of any
planet in a disk with some specified gas disk by looking up that
simulation with the corresponding value of $\gamma$. However, we
caution that the acceleration also has dependence on the sound speed
in the transonic regime, which can break the $\mp\rhog$ degeneracy,
especially if the disk surface density and scale height are set as
free parameters (see \S\ref{sect:disk} for details). 

\begin{figure}[htb]
\centerline{\includegraphics[width=7.0in]{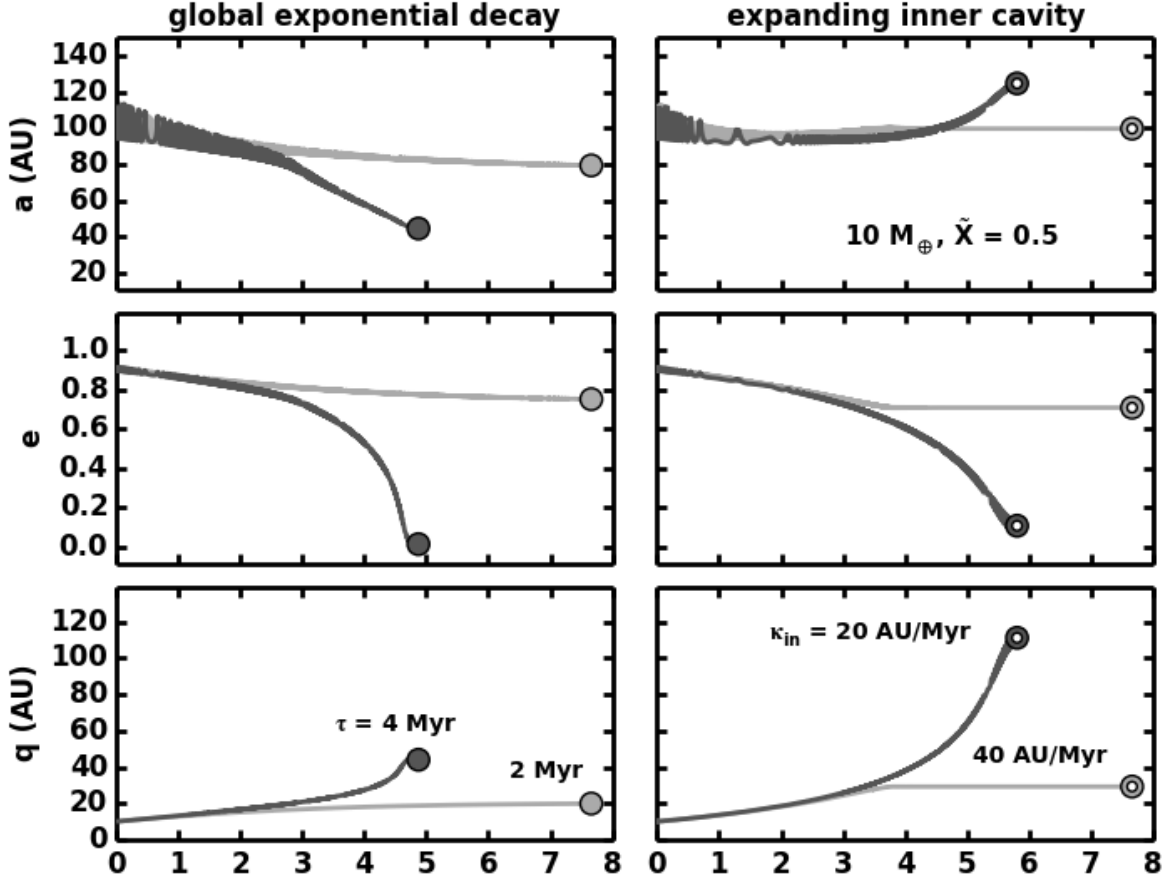}}
\caption{\label{fig:orblf} Evolution of orbits in disks with varying
  gas dispersal modes and time scales.  Orbits are shown for a
  $10\ \Mearth$ planet in disks with $\XX = 0.5$.  Disks evolve either
  by an exponential decay in density (left panels) or with an
  expanding inner \hole\ (right panels). The time scales for dispersal
  are 2 and 4~Myr in the cases of exponential decay. The growth
  rate is 20 and 40~AU/Myr for disks with an expanding inner cavity.
  The darker shaded curves designate a longer-lived disk.  A
  comparison within each panel shows that longer-lived disks cause
  more orbital evolution; a comparison between left and right panels
  illustrates that damping is more effective with a globally decaying
  disk, but that planets settle at larger orbital distances from their
  host star in the case of an expanding inner cavity. The planet in a
  disk with a rapidly growing cavity (light shaded curve and doughnut
  shaped symbols) experienced orbital evolution until the inner
  edge of the disk expanded beyond it, at about 4~Myr. With the
  more slowly expanding inner edge, the planet was able to circularize. 
}
\end{figure}

Aside from the masses of the disk and planet, the time evolution
of the disk surface density also sets the damping time. 
Figure~\ref{fig:orblf} compares results for 10~\mearth\ planets
embedded in $\XX$ = 0.5 disks with different modes and time scales
for disk depletion.  Disk lifetime is clearly important.
Short-lived disks ($\tau = 2$~Myr) or disks with a rapidly expanding 
inner edge ($\openrate = 40$~AU/Myr) are less efficient at damping 
planetary orbits than longer-lived disks. In short-lived disks, 
the acceleration from dynamical friction and gas drag decline too 
rapidly relative to longer-lived disks. Thus, the eccentricity 
evolution in a short-lived disk is limited. 

In disks with an expanding inner cavity, there are two competing
effects. At small $a$, the increasingly large inner disk radius limits
damping and delays circularization compared to disks with a fixed
inner radius.  At large $a$, the relatively static density profile
enables additional damping compared to the exponential decay
models. Combined, these two features of the evolution conspire to
produce circular orbits at larger $a$ than in the exponential decay
models.

When the size of the inner cavity expands rapidly, the inner edge 
of the disk often passes by the planet. This evolution freezes the
orbital elements well before the orbit circularizes.

Overall, the details of disk dissipation clearly have a large impact
on the fate of scattered planets. In disks decaying exponentially on
long time scales, massive planets achieve circular orbits at small
$a$. When disks have slowly growing inner cavities, massive planets
achieve circular orbits at much larger $a$. 
Even lower mass planets whose periastron distance grows only modestly
by virtue of eccentricity damping can have a growing semimajor axis
(e.g., upper right panel in Fig.~\ref{fig:orblf}).  Thus, measuring
the orbital elements for large ensembles of planets with $a \approx$
20--100~AU might provide some insight into the time evolution of the
disk surface density.

\subsection{Simulation results and the final orbital configurations}

\begin{figure}[htb]
\centerline{\includegraphics[width=7.0in]{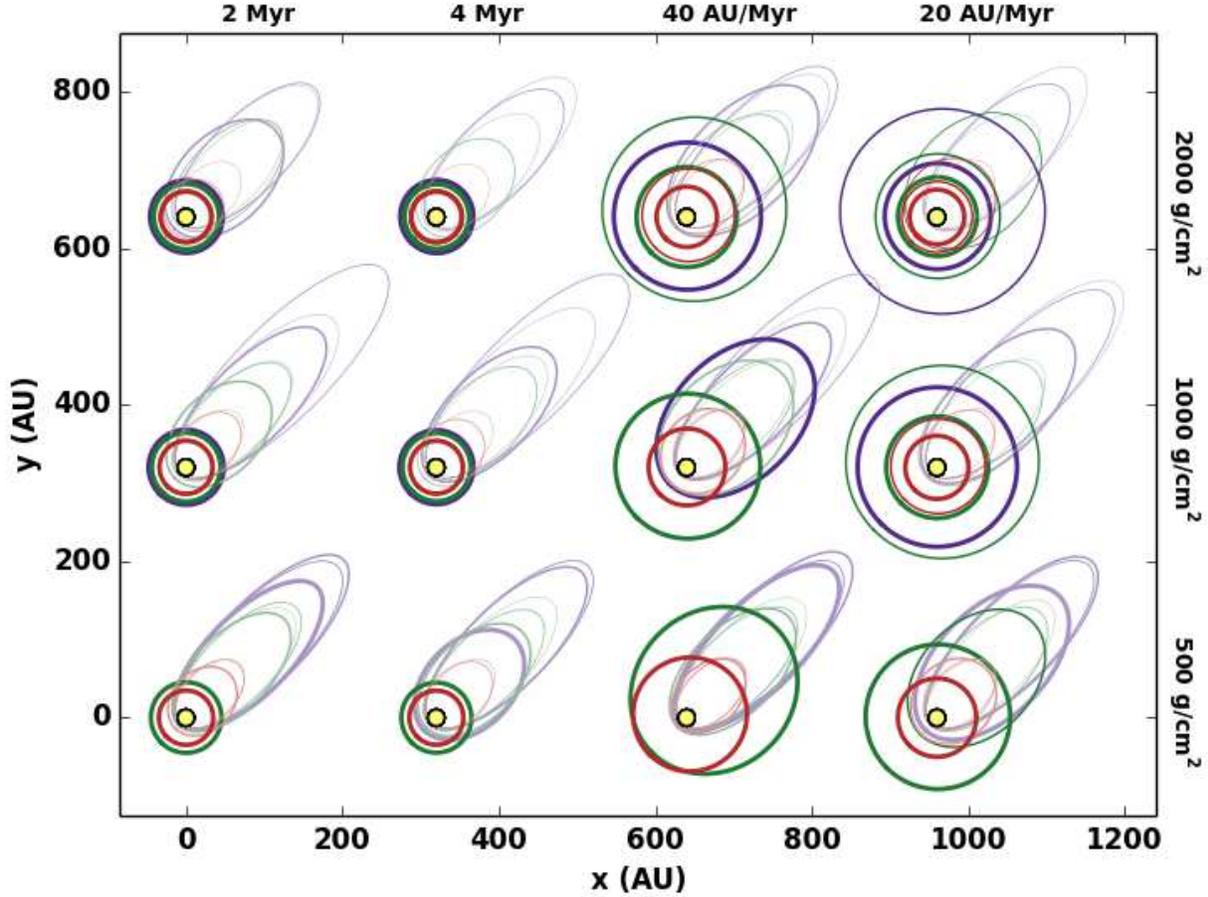}}
\caption{\label{fig:orbs}Simulation outcomes depicted as elliptical
  orbital paths for a suite of models. The $x$-$y$ coordinates give
  planetary positions in their orbital plane; models are offset from
  one another so that they are sorted in rows and columns according to
  model parameters.  The first and second columns correspond to
  exponentially decaying disks with $\tau = 2$ and 4~Myr,
  respectively. The third and fourth columns have disks with expanding
  inner edges with rates of $\openrate = 40$ and 20~AU/Myr, with the
  more slowly expanding gap on the far right.  In each diagram the
  line weight corresponds to planet mass (1, 3, 10, and 30~$\Mearth$),
  while the color indicates the initial apoastron distance (100~AU is
  red, 200~AU is green and 300~AU is blue).  Only the heavily shaded
  trajectories have settled to periastron distances beyond Neptune's
  orbit.}
\end{figure}

Figure \ref{fig:orbs} illustrates the outcomes of the
simulations, showing final orbits for each disk configuration
distinguished by surface density (increasing with vertical position of
the central stars in the plot) and disk evolution mode (exponential
decay on the left and expanding inner \hole\ disk on the right). A
comparison between individual panels illustrates that the outcomes
depend on how effectively the gas can act on a planet before the gas
vanishes.

In the left two columns, the graphic emphasizes the point that planets
tend to circularize relatively close, $a \approx 40$~AU, to the host star. 
Damping depends on the disk lifetime: planets circularize more easily
in long-lived disks than in short-lived disks. Independent of the 
disk lifetime, massive planets circularize more easily than low mass 
planets. For the disks in this study, 10--30~\mearth\ planets achieve 
circular orbits; 1--5~\mearth\ planets remain on eccentric orbits.

In the right two columns, the diagram illustrates how circularization
depends on the mode of disk dispersal. Expanding inner cavities tend 
to leave planets on orbits with large $a$. Disks with slowly expanding
inner cavities have more time to circularize planetary orbits than 
disks with rapidly expanding cavities.  As with exponentially
decaying disks, 10--30~$\Mearth$ planets circularize much more frequently
than smaller mass planets. 

Overall, disks with expanding inner cavities yield planets with a 
broader range of $a$ and $e$ than the exponentially decaying disks.
In the cavity models, varying the disk mass and the expansion time
provide a larger range of circularization time scales compared to
the exponentially decaying disks. Thus, exponentially decaying disks
circularize a few massive planets and leave the rest on orbits with
elements similar to their initial elements. Disks with expanding 
cavities have time to fill the space between circular and high $e$ 
orbits.

\begin{figure}[htb]
\centerline{\includegraphics[width=7.0in]{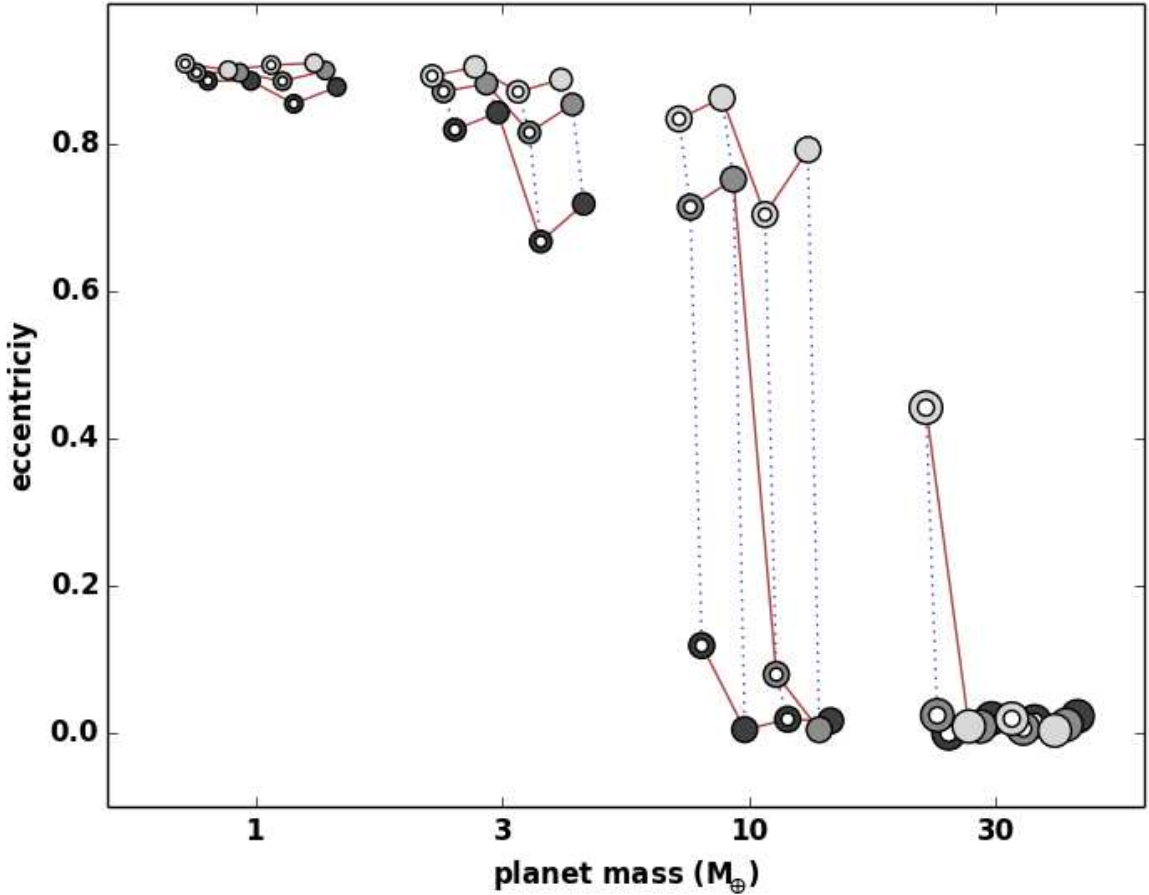}}
\caption{\label{fig:evm} The final eccentricity of planets with
  various masses, reflecting dependence on disk mass and the mode of
  disk depletion---exponential decay or an expanding inner \hole.  The
  points are grouped along the horizontal direction by planet mass (1,
  3, 10 and 30~$\Mearth$; symbol size increases with mass), with slight
  horizontal offsets to distinguish disk mass and depletion mode. The
  filled circles have disks that decay exponentially in time ($\tau =
  2$ and 4~Myr), while the doughnut-shape symbols represent disk
  models with an expanding inner cavity ($\openrate = 20$ and
  40~AU/Myr); Models toward the left have faster depletion times.  In
  all cases, the planets are initially on orbits with a semimajor axis
  of 100~AU and an eccentricity of 0.8; the model with a small planet,
  low disk mass and a rapidly expanding \hole\ (upper left-most
  symbol) experiences little orbital evolution. As planet mass
  increases, the eccentricity damping is more effective; all of the
  $30\ \Mearth$ Neptune analogs (right-most group of symbols) circularize.
}  
\end{figure}

Figure \ref{fig:evm} focuses specifically on eccentricity damping.  It
shows the final eccentricity for models grouped by planet mass.  As in
the previous figures, the dependence is clear: planets with larger
mass interact more strongly, with strong eccentricity
damping for the most massive planets, the Neptune analogs in our
runs. The Earth-mass planets experience comparatively little damping.
Planets with masses between these two limits have eccentricities that
are most sensitive to the details of the disk and its evolution.

Making other choices for $\ain$ and $\aout$ leads to qualitatively
similar results. When $\ain < q = a (1 - e)$, circularization occurs
earlier and at smaller orbital distance. This effect is strongest in
exponentially decaying disks and for the most massive planets. For
example, in an exponentially decaying disk ($\XX = 0.5$, $\tau =
4$~Myr) with the inner edge moved to $\ain$ = 5~AU, a 30~$\Mearth$
planet with an initial apoastron distance of 200~AU ($a = 105$~AU, $e
= 0.9$) ends up on a circular orbit at $a \approx q \approx 30$~AU,
compared to 40~AU in our baseline case with the more distant inner
disk edge.

Extending the outer edge of the disk past the initial apoastron
distance has the opposite effect. In our baseline simulations one set
of scattered planets has an initial apoastron distance of
$a(1+e)=300$~AU, past the edge of the disk at $\aout=200$~AU.
Extending the disk edge to 350~AU causes orbits to settle at larger
final periastron distances; the more massive planets can circularize
at greater orbital distances. Because the gas density is lower at
larger distances, the effect is significant only if the inner edge is
expanding. 

Details of disk dispersal also matter. In an extreme case, a
30~$\Mearth$ planet achieves a circularized orbit at $a \approx q
\approx 150$~AU in a low-mass, extended disk with a slowly expanding
inner edge ($\XX=0.25$, $\aout=350$~AU, and $\openrate=20$~AU/yr).
For comparison, the same planet in a similar disk with an inner edge
at 200~AU circularizes around $a \approx q \approx 100$~AU.

\subsection{Summary of simulation results}

Our suite of simulations shows a variety of outcomes, with several clear 
trends.

\begin{itemize}
\item Circularization of eccentric orbits is most effective for
  massive planets, which create the strongest wakes in the gas
  disk. The orbits of Earth-mass or smaller planets tend to remain
  eccentric. More massive super-Earths circularize well within the
  lifetime of a disk, and may initiate Type I radial migration
  and/or continue to grow by gas accretion. 

\item Disks with exponential decay in surface density generally damp
  the orbits of scattered planets, drawing them inward to smaller
  orbital distances. Only the more massive super-Earths and Neptunes
  circularize. All low eccentricity orbits reside inside 50~AU.
  Smaller planets can achieve moderately eccentric orbits at larger
  orbital distances.

\item Disks with an expanding inner \hole\ have a broader range of
  outcomes. Super-Earths can circularize beyond~100~AU, depending
  on when the expanding inner edge of the disk overtakes their orbital
  distance. Smaller planets can be pushed outward to orbital distances
  in excess of 200~AU, although their eccentricity remains high.

\end{itemize}
Additional simulations test other aspects of this model. For example,
by adjusting the surface density profile index $p$
(eq.~[\ref{eq:disk}]), we find that a steeper (shallower) disk surface
density profile tends to circularize planets less (more) efficiently
for a given $\Soh$ and at smaller (larger) orbital
distances. 

\section{Discussion}
\label{sect:discuss}

We investigate a scenario that allows scattered planets to acquire circular 
orbits in remote regions of planetary systems through interaction with
an extended gas disk. This work is motivated by earlier simulations of
gas giant formation \citep{bk11a}, in which super-Earths (failed gas
giant cores) are scattered onto stable, eccentric orbits at large
semimajor axes. Here, we use a simple parameterization of the
acceleration that a planet experiences from dynamical friction and
aerodynamical drag, and we calculate orbits around a Sun-like star
with an evolving gas disk. Our focus is on eccentricity damping, not
on any radial migration from differential torques
\citep{lin79,gt79,war97} that might occur after the planet settles
onto a circular orbit.

Our numerical models show that the final orbits of scattered planets
depend primarily on their mass. For disks on the high-mass end of the 
distribution observed in T Tauri disks \citep[e.g.,][]{and13}, 
planets more massive than the Earth experience substantial orbital 
evolution. Neptunes damp efficiently.  Earth mass planets damp little.
Because eccentricity and inclination are generally correlated, the
models yield clear correlations between planet mass and the orbital
elements $e$ and $i$.

Precise predictions for orbital architectures accessible with direct 
imaging depend on how the circumstellar disk vanishes. If mass simply 
decays exponentially with time everywhere in the disk, our models
predict super-Earth and Neptune mass planets on low $e$ orbits close 
to the host star. However, current data suggest that the transition
disks have expanding inner cavities \citep{rib14}. In these disks,
we expect massive planets on roughly circular orbits at larger 
distances from the host star.

These correlations between the planet mass and the final semimajor axis
and eccentricity are much different from migration models, where planets 
remain on circular orbits unless perturbed by another nearby planet 
\citep{war97}.  Models for {\it in situ} formation also leave massive 
planets on fairly circular orbits \citep{helled2013}.  Identification 
of the trends predicted by the simulations would help to distinguish 
scattering from {\it in situ} formation and Type I radial migration.

The frequency of remote super-Earths---put in place according to our
scenario---depends on the prevalence of massive circumstellar disks.  
Our models require a relatively massive disk, with $\Soh\ge
500$~g/cm$^2$ at 1~AU. With a power-law index of $p=1$, our models have
total disk mass in excess of 0.07~$\Msolar$.  Observations of young
stellar systems show a wide range of disk masses and configurations
\citep[e.g.,][]{pie07,and11,den13,pie14}. Thus, remote super-Earths may be
possible around only $\lae 10$\% of stars \citep[cf.][]{and13,naj14}.
Fortunately, these same massive disks are the most likely to produce
multiple super-Earths and the larger planets needed to scatter them
\citep{bk11a}. Neptune analogs or more massive scattered planets may
be able to settle in even less massive disks. We will explore this
possibility in future work.

Imaging observations and planned surveys with Gemini Planet Imager
\citep{mac14}, Subaru \citep{tamu14} and other facilities are beginning
to map out the frequency of distant planets around their host stars.  
The detections are presently limited to Jupiter-size objects beyond
roughly 10~AU. Extending these observations to lower mass planets 
with larger semimajor axes may yield tests of our models.

The planets around HR~8799 \citep{mar08} show the promise of imaging
surveys. This planetary system has four super-Jupiters likely on low
eccentricity orbits \citep{cur12b}, with the most distant at roughly
70~AU from its 1.5~$\Msolar$ host star.  The masses of these planets
are much greater than in our models, but if they follow the trend of
rapid circularization with increasing mass, then processes described
here may have been at work in damping the outer planets to
their observed orbital configuration. 

Planet-disk interactions may also have contributed to the dynamics of
the early solar system.  Beyond the orbit of Neptune, icy objects such
as Sedna \citep{bro04} are too small to interact with a massive
gaseous disk.  However, correlations in the orbital parameters of
Sedna and similar objects have led to speculation that a super-Earth
with mass 2--15~$\Mearth$ resides on a low-eccentricity orbit between
200 AU and 300 AU \citep[][]{tru14}.  If this planet exists, {\it in
  situ} formation \citep{kb04,stern2005} and migration from inside
30~AU \citep{morbi2013} seem unlikely.  If the planet is massive and 
interacted with a massive, extended disk that dispersed from the inside 
out, formation at small $a$ followed by scattering is plausible.
Discovering this planet -- barely below the current
threshold of detectability \citep[][]{tru14} -- would provide an
excellent test of our scenario and might give us a new probe into the
structure and evolution of the Sun's circumstellar disk.

\acknowledgements

We are grateful to M.\ Geller for comments and advice on presentation.
We also appreciate the thoughtful comments of an anonymous referee.
We acknowledge NASA for a generous allotment of computer time on 
the NCCS 'discover' cluster. Portions of this project were supported 
by the {\it NASA Astrophysics Theory} and {\it Origins of Solar Systems} 
programs through grant NNX10AF35G, and the {\it NASA Outer Planets Program} 
through grant NNX11AM37G.

%% Figures.....

\begin{thebibliography}{99}

\bibitem[Adachi et al.(1976)]{ada76} Adachi, I., Hayashi, C., 
\& Nakazawa, K.\ 1976, Progress of Theoretical Physics, 56, 1756 

\bibitem[Adams et al.(2009)]{adac09} Adams, F.~C., Cai, M.~J., 
\& Lizano, S.\ 2009, \apjl, 702, L182 

%\bibitem[Alexander \& Armitage(2009)]{ale09} Alexander, R.~D., \&
%Armitage, P.~J.\ 2009, \apj, 704, 989

\bibitem[Alexander et al.(2013)]{ale13} Alexander, R., 
Pascucci, I., Andrews, S., Armitage, P., 
\& Cieza, L.\ 2013, 
Protostars and Planets VI (in press; arXiv:1311.1819)

\bibitem[Andrews et al.(2011)]{and11} Andrews, S.~M., Wilner, 
D.~J., Espaillat, C., et al.\ 2011, \apj, 732, 42 

\bibitem[Andrews et al.(2012)]{and12} Andrews, S.~M., Wilner, 
D.~J., Hughes, A.~M., et al.\ 2012, \apj, 744, 162 

\bibitem[Andrews et al.(2013)]{and13} Andrews, S.~M., 
Rosenfeld, K.~A., Kraus, A.~L., \& Wilner, D.~J.\ 2013, \apj, 771, 129 

\bibitem[Andrews \& Williams(2005)]{and05} Andrews, S.~M., \&
  Williams, J.~P.\ 2005, \apj, 631, 1134

% Nice but with 5 planets.
\bibitem[Batygin et al.(2012)]{bat12} Batygin, K., Brown, 
M.~E., \& Betts, H.\ 2012, \apjl, 744, L3 

\bibitem[Beust et al.(2014)]{beu14} Beust, H., Augereau,
  J.-C., Bonsor, A., et al.\ 2014, \aap, 561, A43

\bibitem[Binney \& Tremaine(1987)]{bintre87} Binney, J., \&
Tremaine, S.\ 1987, Galactic Dynamics (Princeton: Princeton University
Press)

\bibitem[Birnstiel \& Andrews(2014)]{bir14} Birnstiel,
  T., \& Andrews, S.~M.\ 2014, \apj, 780, 153

\bibitem[Bromley \& Kenyon(2006)]{bk06} 
Bromley, B., \& Kenyon, S. J.
2006, \aj, 131, 2737

\bibitem[Bromley \& Kenyon(2011a)]{bk11a} 
Bromley, B.~C., \& Kenyon, S.~J.\ 2011a, \apj, 731, 101 

\bibitem[Bromley \& Kenyon(2011b)]{bk11b} 
Bromley, B.~C., \& Kenyon, S.~J.\ 2011b, \apj, 735, 29

% saturn ring
\bibitem[Bromley \& Kenyon(2013)]{bk13} Bromley, B.~C.,
  \& Kenyon, S.~J.\ 2013, \apj, 764, 192

% Sedna discovery
\bibitem[Brown et al.(2004)]{bro04} Brown, M.~E., Trujillo, 
C., \& Rabinowitz, D.\ 2004, \apj, 617, 645 

\bibitem[Calvet et al.(2005)]{cal05} Calvet, N., D'Alessio, 
P., Watson, D.~M., et al.\ 2005, \apjl, 630, L185 

\bibitem[Chambers(2014)]{cha14} Chambers, J.~E.\ 2014, 
\icarus, 233, 83 

\bibitem[Chatterjee et al.(2008)]{cha08} Chatterjee, S., 
Ford, E.~B., Matsumura, S., \& Rasio, F.~A.\ 2008, \apj, 686, 580 

\bibitem[Chiang \& Goldreich(1997)]{cha97} 
Chiang, E.~I., \& Goldreich, P.\ 1997, \apj, 490, 368 

\bibitem[Clarke et al.(2001)]{cla01} Clarke, C.~J., Gendrin, 
A., \& Sotomayor, M.\ 2001, \mnras, 328, 485 

% outward migration of gas giants
\bibitem[Crida et al.(2009)]{cri09} Crida, A., Masset, F., 
\& Morbidelli, A.\ 2009, \apjl, 705, L148 


% fomalhaut b
\bibitem[Currie et al.(2012a)]{cur12a} Currie, T., Debes, J., 
Rodigas, T.~J., et al.\ 2012a, \apjl, 760, L32 

\bibitem[Currie et al.(2012b)]{cur12b} Currie, T., Fukagawa, 
M., Thalmann, C., Matsumura, S., \& Plavchan, P.\ 2012b, \apjl, 755, L34 

\bibitem[Currie et al.(2008)]{cur08} Currie, T., Kenyon, 
S.~J., Balog, Z., et al.\ 2008, \apj, 672, 558 


\bibitem[D'Angelo et al.(2011)]{dan11} D'Angelo, G., Durisen, R.~H.,
  \& Lissauer, J.~J.\ 2011, Exoplanets, edited by S.~Seager.~ Tucson,
  AZ: University of Arizona Press, 2011, 526 pp.~ ISBN
  978-0-8165-2945-2., p.319-346, 319

% GASPS herschel survey of gas and dust...
\bibitem[Dent et al.(2013)]{den13} Dent, W.~R.~F., Thi, 
W.~F., Kamp, I., et al.\ 2013, \pasp, 125, 477 

\bibitem[Dittrich et al.(2013)]{dit13} Dittrich, K., Klahr, 
H., \& Johansen, A.\ 2013, \apj, 763, 117 

\bibitem[Dokuchaev(1964)]{dok64} Dokuchaev, V.~P.\ 1964, 
\sovast, 8, 23 

\bibitem[Duncan et al.(1998)]{dun98} Duncan, M.~J., Levison, 
H.~F., \& Lee, M.~H.\ 1998, \aj, 116, 2067 


% planet-planet scattering...e=0.8.???
\bibitem[Ford \& Rasio(2008)]{for08} Ford, E.~B., \&
  Rasio, F.~A.\ 2008, \apj, 686, 621


% HD 95086 b with the Gemini Planet Imager
\bibitem[Galicher et al.(2014)]{2014A&A...565L...4G} Galicher, R.,
  Rameau, J., Bonnefoy, M., et al.\ 2014, \aap, 565, L4


\bibitem[Goldreich \& Tremaine(1979)]{gt79} Goldreich, P., \&
Tremaine, S.\ 1979, \apj, 233, 857


\bibitem[Guilet et al.(2013)]{gui13} Guilet, J.,
  Baruteau, C., \& Papaloizou, J.~C.~B.\ 2013, \mnras, 430, 1764

\bibitem[Hahn \& Malhotra(1999)]{hah99} Hahn, J.~M., \& Malhotra, R.\ 1999, \aj, 117, 3041 

\bibitem[Haisch et al.(2001)]{hai01} Haisch, K.~E., Jr., 
Lada, E.~A., \& Lada, C.~J.\ 2001, \apjl, 553, L153 

\bibitem[Helled et al.(2013)]{helled2013} Helled, R., Bodenheimer, 
P., Podolak, M., et al.\ 2013, arXiv:1311.1142 

\bibitem[Hori \& Ikoma(2011)]{hor11} Hori, Y., \& Ikoma, M.\ 2011,
  \mnras, 416, 1419

\bibitem[Hoyle \& Lyttleton(1941)]{hoy41} Hoyle, F., \& Lyttleton,
  R.~A.\ 1941, \mnras, 101, 227

\bibitem[Johansen et al.(2007)]{joh07} Johansen, A., Oishi, 
J.~S., Mac Low, M.-M., et al.\ 2007, \nat, 448, 1022 

\bibitem[Juri{\'c} 
\& Tremaine(2008)]{jur08} Juri{\'c}, M., \& Tremaine, S.\ 2008, \apj, 686, 603 


% discovery of fomalhaut b
\bibitem[Kalas et al.(2008)]{kal08} Kalas, P., Graham, J.~R., 
Chiang, E., et al.\ 2008, Science, 322, 1345 

\bibitem[Kalas et al.(2013)]{kal13} Kalas, P., Graham, J.~R., 
Fitzgerald, M.~P., \& Clampin, M.\ 2013, \apj, 775, 56 

\bibitem[Keane et al.(2014)]{kea14} Keane, J.~T., Pascucci, 
I., Espaillat, C., et al.\ 2014, arXiv:1404.0709 

\bibitem[Kennedy \& Kenyon(2008)]{kk08} Kennedy, G.~M., \& Kenyon,
S.~J.\ 2008, \apj, 673, 502

\bibitem[Kenyon \& Bromley(2001)]{kb01}
Kenyon, S.~J., \& Bromley, B.~C.\ 2001, \aj, 121, 538 

\bibitem[Kenyon \& Bromley(2002)]{kb02} Kenyon, S. J., \& Bromley, B. C.
2002a, AJ, 123, 1757

\bibitem[Kenyon \& Bromley(2004)]{kb04} Kenyon, S.~J., \& Bromley,
B.~C.\ 2004, \aj, 127, 513

% oligarch2chaotic
\bibitem[Kenyon \& Bromley(2006)]{kb06} Kenyon, S.~J., \& Bromley,
B.~C.\ 2006, \aj, 131, 1837

\bibitem[Kenyon \& Bromley(2009)]{kb09} Kenyon, S.~J., \& Bromley,
B.~C.\ 2009, \apjl, 690, L140

\bibitem[Kenyon \& Bromley(2010)]{kb10} Kenyon, S.~J., \& Bromley,
B.~C.\ 2010, \apjs, 188, 242

\bibitem[Kenyon \& Bromley(2014)]{kb14a} Kenyon, S.~J.,
  \& Bromley, B.~C.\ 2014, \aj, 147, 8

% fomalhaut b
\bibitem[Kenyon et al.(2014)]{kcb14} Kenyon, S.~J., Currie, 
T., \& Bromley, B.~C.\ 2014, \apj, 786, 70 

\bibitem[Kenyon \& Luu(1998)]{kl98} Kenyon, S.~J., \& Luu,
  J.~X.\ 1998, \aj, 115, 2136

% core formation through agglomeration
\bibitem[Kobayashi et al.(2011)]{kob11} Kobayashi, H.,
  Tanaka, H., \& Krivov, A.~V.\ 2011, \apj, 738, 35

\bibitem[Lambrechts \& Johansen(2012)]{lam12}
  Lambrechts, M., \& Johansen, A.\ 2012, \aap, 544, A32


\bibitem[Lee \& Stahler(2014)]{lee14} Lee, A.~T., \&
  Stahler, S.~W.\ 2014, \aap, 561, A84

\bibitem[Levison et al.(1998)]{lev98} Levison, H.~F., 
Lissauer, J.~J., \& Duncan, M.~J.\ 1998, \aj, 116, 1998 

\bibitem[Lin \& Bodenheimer(1982)]{lin82} Lin, D.~N.~C.,
  \& Bodenheimer, P.\ 1982, \apj, 262, 768


\bibitem[Lin \& Papaloizou(1979)]{lin79} Lin, D.~N.~C., \& Papaloizou,
J.\ 1979, \mnras, 186, 799

\bibitem[Lin \& Papaloizou(1980)]{lin80} Lin, D.~N.~C.,
  \& Papaloizou, J.\ 1980, \mnras, 191, 37

\bibitem[Lovelace et al.(2008)]{lov08} Lovelace, R.~V.~E., 
Romanova, M.~M., \& Barnard, A.~W.\ 2008, \mnras, 389, 1233 


\bibitem[Lynden-Bell \& Pringle(1974)]{lyn74}
  Lynden-Bell, D., \& Pringle, J.~E.\ 1974, \mnras, 168, 603


\bibitem[Macintosh et al.(2014)]{mac14} Macintosh, B., 
Graham, J.~R., Ingraham, P., et al.\ 2014, arXiv:1403.7520 

\bibitem[Marois et al.(2008)]{mar08} Marois, C., Macintosh, 
B., Barman, T., et al.\ 2008, Science, 322, 1348 

\bibitem[Morbidelli (2013)]{morbi2013} Morbidelli, A.\ 2013, 
Planets, Stars and Stellar Systems.~Volume 3: Solar and Stellar Planetary 
Systems, 63 

\bibitem[Najita \& Kenyon (2014)]{naj14} Najita, J. R., \& Kenyon, S. J.\ 2014,
\mnras, submitted

\bibitem[Ohtsuki et al.(1988)]{oht88} Ohtsuki, K., Nakagawa, 
Y., \& Nakazawa, K.\ 1988, \icarus, 75, 552 

\bibitem[Ostriker(1999)]{ost99} Ostriker, E.~C.\ 1999, \apj, 
513, 252 

\bibitem[Owen et al.(2012)]{owe12} Owen, J.~E., Clarke, 
C.~J., \& Ercolano, B.\ 2012, \mnras, 422, 1880 


\bibitem[Papaloizou et al.(2007)]{pap07} Papaloizou, 
J.~C.~B., Nelson, R.~P., Kley, W., Masset, F.~S., 
\& Artymowicz, P.\ 2007, Protostars and Planets V, 655 

\bibitem[Papaloizou \& Terquem(1999)]{pap99} Papaloizou,
  J.~C.~B., \& Terquem, C.\ 1999, \apj, 521, 823

\bibitem[Paardekooper(2009)]{paa09} Paardekooper, S.-J.\ 2009, \aap,
  506, L9

% dust disk structure
\bibitem[Pascucci et al.(2004)]{pas04} Pascucci, I., Wolf, S.,
  Steinacker, J., et al.\ 2004, \aap, 417, 793

% Probing the structure of protoplanetary disks: 
% a comparative study of DM Tau, LkCa 15, and MWC 480
\bibitem[Pi{\'e}tu et 
al.(2007)]{pie07} Pi{\'e}tu, V., Dutrey, A., \& Guilloteau, S.\ 2007, \aap, 467, 163 

\bibitem[Pi{\'e}tu et al.(2014)]{pie14} Pi{\'e}tu, V.,
  Guilloteau, S., Di Folco, E., Dutrey, A., \& Boehler, Y.\ 2014,
  \aap, 564, A95

% min core mass for accretion
\bibitem[Piso \& Youdin(2014)]{pis14} Piso, A.-M.~A., \& Youdin,
  A.~N.\ 2014, \apj, 786, 21

\bibitem[Pollack et al.(1996)]{pol96} Pollack, J.~B., 
Hubickyj, O., Bodenheimer, P., Lissauer, J.~J., Podolak, M., 
\& Greenzweig, Y.\ 1996, \icarus, 124, 62 

\bibitem[Rafikov(2011)]{raf11} Rafikov, R.~R.\ 2011, \apj, 
727, 86 

\bibitem[Rasio \& Ford (1996)]{rasio1996} Rasio, F.~A., \& Ford, E.~B.\ 1996, Science, 274, 954 

% outward migration of neptunes
\bibitem[Raymond \& Bonsor(2014)]{ray14} Raymond, S.~N.,
  \& Bonsor, A.\ 2014, \mnras, 442, L18


\bibitem[Ribas et al.(2014)]{rib14} Ribas, {\'A}., Mer{\'{\i}}n, B.,
  Bouy, H., \& Maud, L.~T.\ 2014, \aap, 561, A54

\bibitem[Ruden(2004)]{rud04} Ruden, S.~P.\ 2004, \apj, 605, 
880 

\bibitem[Ruderman \& Spiegel(1971)]{rud71} Ruderman, 
M.~A., \& Spiegel, E.~A.\ 1971, \apj, 165, 1 

\bibitem[Ruffert(1996)]{ruf96} Ruffert, M.\ 1996, \aap, 311, 817 

\bibitem[Shima et al.(1985)]{shi85} Shima, E., Matsuda, T., 
Takeda, H., \& Sawada, K.\ 1985, \mnras, 217, 367 

\bibitem[Stern(2005)]{stern2005} Stern, S.~A.\ 2005, \aj, 129, 
526 

%\bibitem[Stewart \& Ida(2000)]{ste00} Stewart, G.~R., \& Ida, S.\
%2000, \icarus, 143, 28

% fom b
\bibitem[Tamayo(2014)]{tamo14} Tamayo, D.\ 2014, \mnras, 438, 
3577 

% seeds subaru
\bibitem[Tamura(2014)]{tamu14} Tamura, M.\ 2014, IAU 
Symposium, 299, 12 

\bibitem[Tanaka et al.(2002)]{tan02} Tanaka, H., Takeuchi, 
T., \& Ward, W.~R.\ 2002, \apj, 565, 1257 

%\bibitem[Taylor(1981)]{1981A&A...103..288T} Taylor, D.~B.\ 1981, \aap, 103, 
%  288  % horseshoe orbit study

%\bibitem[Thommes et al.(2003)]{tho03} Thommes, E.~W., Duncan, 
%M.~J., \& Levison, H.~F.\ 2003, \icarus, 161, 431 

\bibitem[Trujillo 
\& Sheppard(2014)]{tru14} Trujillo, C.~A., \& Sheppard, S.~S.\ 2014, \nat, 507, 471 

\bibitem[Tsiganis et al.(2005)]{tsi05} Tsiganis, K., Gomes, 
R., Morbidelli, A., \& Levison, H.~F.\ 2005, \nat, 435, 459 

\bibitem[Ward(1997)]{war97} Ward, W.~R.\ 1997, \icarus, 126, 
261 

\bibitem[Weidenschilling(1977)]{wei77} Weidenschilling, 
S.~J.\ 1977, \mnras, 180, 57 

%\bibitem[Weidenschilling(2000)]{wei00} Weidenschilling, 
%S.~J.\ 2000, \ssr, 92, 295 

%\bibitem[Wetherill \& Stewart(1989)]{wet89} Wetherill, G.~W., \&
%Stewart, G.~R.\ 1989, \icarus, 77, 330

%\bibitem[Whipple(1973)]{whi73} Whipple, F.~L.\ 1973, NASA 
%Special Publication, 319, 355 

\bibitem[Youdin \& Kenyon(2013)]{you13} Youdin, A.~N.,
  \& Kenyon, S.~J.\ 2013, Planets, Stars and Stellar Systems.~Volume
  3: Solar and Stellar Planetary Systems, 1


\end{thebibliography}
\end{document}